\documentclass[aps,unsortedaddress,showpacs,amsmath]{revtex4}

\begin{document}

\title{Vortex tubes in velocity fields of laboratory isotropic turbulence: dependence on the Reynolds number}

\author{Hideaki Mouri}
\email{hmouri@mri-jma.go.jp}

\author{Akihiro Hori}
\altaffiliation[Also at ]{Japan Weather Association, Higashi-Ikebukuro 3-1-1, Toshima, Tokyo 170-6055, Japan.}

\author{Yoshihide Kawashima}
\altaffiliation[Also at ]{Japan Weather Association, Higashi-Ikebukuro 3-1-1, Toshima, Tokyo 170-6055, Japan.}
\affiliation{Meteorological Research Institute, Nagamine 1-1, Tsukuba 305-0052, Japan}

\date{\today}

\begin{abstract}
The streamwise and transverse velocities are measured simultaneously in isotropic grid turbulence at relatively high Reynolds numbers, Re$_{\lambda} \simeq 110$--330. Using a conditional averaging technique, we extract typical intermittency patterns, which are consistent with velocity profiles of a model for a vortex tube, i.e., Burgers vortex. The radii of the vortex tubes are several of the Kolmogorov length regardless of the Reynolds number. Using the distribution of an interval between successive enhancements of a small-scale velocity increment, we study the spatial distribution of vortex tubes. The vortex tubes tend to cluster together. This tendency is increasingly significant with the Reynolds number. Using statistics of velocity increments, we also study the energetical importance of vortex tubes as a function of the scale. The vortex tubes are important over the background flow at small scales especially below the Taylor microscale. At a fixed scale, the importance is increasingly significant with the Reynolds number. 
\end{abstract}

%\keywords{Suggested keywords}%Use showkeys class option if keyword display desired

\maketitle

\section{INTRODUCTION}
\label{s1}

By using direct numerical simulations \cite{vm91,j93,tmi99} and bubble/cavitation visualization experiments \cite{d91,cdc95,lvmb00}, it has been established that turbulence contains vortex tubes. Regions of intense vorticity are organized into tubes. Their radii and lengths are, respectively, of the orders of the Kolmogorov length $\eta$ and the integral length $L$. Their lifetimes are several turnover times of the largest eddies. The vortex tubes occupy a small fraction of the volume and are embedded in a background flow that is random and of large scales (see also reviews \cite{f95,sa97,s99}). Here we study vortex tubes in velocity fields of experimental turbulence as a function of the Reynolds number Re$_{\lambda}$.

Especially when the Reynolds number is high, effects of vortex tubes on the velocity field are of interest. The velocity signal at small scales is enhanced only in a fraction of the volume \cite{bt49}. This small-scale intermittency is attributable to vortex tubes \cite{f95,sa97,s99,n97,cg99}. With increasing the Reynolds number, turbulence becomes more intermittent \cite{sa97}.

Large values of the Reynolds number, Re$_{\lambda} \gg 100$, are achieved only in experiments, where a measurement is made with a probe suspended in the flow and a one-dimensional cut of the velocity field is obtained. The measurement often deals with the velocity component in the mean-flow direction alone (hereafter, the streamwise velocity $u$), but the transverse velocity $v$ is more suited to detecting circulation flows such as those associated with vortex tubes \cite{n97,cg99,mtk99}. A simultaneous measurement of the two velocity components is desirable.

For laboratory experiments, there are several possible configurations. Among them, a flow downstream of a turbulence-generating grid is preferable. This grid turbulence is isotropic and free from external influences such as mean shear, and hence is suited to isolating effects of vortex tubes at each of the scales. Although the Reynolds number is not very high and the inertial subrange is almost absent, grid turbulence is a natural step from direct numerical simulations of isotropic turbulence with small Re$_{\lambda}$ values to laboratory or field experiments of anisotropic turbulence with very large Re$_{\lambda}$ values. 

We carried out simultaneous measurements of the streamwise and transverse velocities in grid turbulence with a range of the Reynolds number, Re$_{\lambda} \simeq 110$--330. Since our wind tunnels as well as our grids were large, the Re$_{\lambda}$ values exceed those in past studies of grid turbulence. The exception is a special but old experiment with high air pressure \cite{kv66}, for which the Reynolds number was in the range Re$_{\lambda} = 260$--670. Thus our data provide a unique opportunity to study vortex tubes in isotropic turbulence with a wide range of the Reynolds number. The preliminary studies had been reported by Mouri et al. \cite{mhk00}, but the present data are new and are analyzed more carefully and extensively.

In Sec.~\ref{s2}, we describe our experiments. In Sec.~\ref{s3}, we present a model for a vortex tube. In Sec.~\ref{s4}, a conditional averaging technique is used to extract typical intermittency patterns from the data. The patterns turn out to be consistent with velocity profiles expected for the model tube. Also, statistics of velocity increments are used to study the energetical importance of vortex tubes as a function of the scale. In Sec.~\ref{s5}, we present our conclusion with remarks for future experimental researches.

\section{EXPERIMENTS}
\label{s2}

The experiments were done in two wind tunnels of Meteorological Research Institute. Their test sections were of 0.8 $\times$ 0.8 $\times$ 3 and 3 $\times$ 2 $\times$ 18 m in size. Turbulence was produced by placing a grid across the entrance to the test section. The grid consisted of two layers of uniformly spaced rods, the axes of which were perpendicular to each other. We used three grids. The separations of the axes of their adjacent rods were $M$ = 0.1, 0.2, and 0.4 m. The cross sections of the rods were, respectively, 0.02 $\times$ 0.02, 0.04 $\times$ 0.04, and 0.06 $\times$ 0.06 m. The mean wind was set to be $U$ $\simeq$ 4, 8, 12, or 16 m~s$^{-1}$.

The streamwise ($U + u$) and transverse ($v$) velocities were measured simultaneously with a hot-wire anemometer. The anemometer was composed of a crossed-wire probe and a constant temperature system. The probe was positioned on the tunnel axis. The hot wires were platinum-coated tungsten filaments, 5 $\mu$m in diameter, 1.25 mm in effective length, 1 mm in separation, 280$^{\circ}$C in temperature, and oriented at $\pm 45 ^{\circ}$ to the mean-flow direction. The calibration was done before and after each of the measurements. 

The signal was low-pass filtered at $f_c$ = 4--16 kHz with 24 dB per octave and sampled digitally at $f_s$ = 8--32 kHz with 16-bit resolution. To avoid aliasing, the sampling frequency was set to be twice of the filter cutoff frequency, $f_s = 2 f_c$. The entire length of the signal was 2 $\times$ 10$^7$ points, which is large enough to guarantee the convergence of the statistics \cite{agha84,mkfnt99}.

For each set of the grid and the mean wind speed, measurements were made at several positions downstream of the grid. We first computed the flatness factors $\langle u^4 \rangle / \langle u^2 \rangle ^2$ and $\langle v^4 \rangle / \langle v^2 \rangle ^2$, where $\langle \cdot \rangle$ denotes an average. The data with the flatness factors being close to the Gaussian value of 3 were used in further analyses because we are to study fully developed turbulence. If the measurement position is too close to or far from the grid, the turbulence is still developing or already decaying, and its flatness factors are different from the Gaussian value \cite{mthk02}. Except for the measurements done in the $0.8 \times 0.8 \times 3$ m tunnel, the flow was almost isotropic. The ratio between the streamwise- and transverse-velocity dispersions $\langle u^2 \rangle / \langle v^2 \rangle$ was close to unity.

The turbulence level, i.e., the ratio of the root-mean-square streamwise fluctuation $\langle u^2 \rangle ^{1/2}$ to the mean streamwise velocity $U$, was always less than 10\%. This good characteristic of grid turbulence allows us to rely on the frozen-eddy hypothesis of Taylor \cite{t38}, $\partial / \partial t $ = $-U \partial / \partial x$, which converts temporal variations into spatial variations in the mean-wind direction.

To obtain the highest spatial resolution, we set $f_c$ in such a way that $U / f_c$ is close to the probe size, $\sim 1$ mm. Since the probe size is greater than the Kolmogorov length, 0.2--0.5 mm, the smallest-scale motions of the flow were filtered out. The present resolution is nevertheless typical of hot-wire anemometry. Although the RELIEF technique achieves a higher resolution \cite{n97}, it is applicable only to the transverse velocity.

%%%% Place Table 1 %%%%

Consequently we obtained 10 sets of turbulence data. The experimental conditions are summarized in Table \ref{t1}. The flow parameters such as the integral length $L$, the Kolmogorov length $\eta$, the Taylor microscale $\lambda$, and the Reynolds number Re$_{\lambda}$ are also summarized there.

\section{MODEL FOR VORTEX TUBE}
\label{s3}

Before describing our experimental results, it is helpful to present a model for a vortex tube, Burgers vortex \cite{b48}. This model is an axisymmetric steady circulation in a strain field. In cylindrical coordinates, the circulation and the strain field are written respectively as
\begin{subequations}
\begin{equation}
\label{eq1a}
  u_{\Theta} \propto \frac{2 \nu}{a r} 
             \left[ 1 - \exp \left( - \frac{a r^2}{4 \nu} \right) \right]
  \quad (a > 0),
\end{equation}
and
\begin{equation}
\label{eq1b}
\left( u_r , u_{\Theta}, u_z \right) = 
\left( - \frac{1}{2} a r, 0, a z\right).
\end{equation}
\end{subequations}
Here $\nu$ is the kinematic viscosity. The above Eq. (\ref{eq1a}) describes a rigid-body rotation for small radii, and a circulation decaying in radius for large radii. The velocity is maximal at $r$ = $r_0$ = $2.24 (\nu / a)^{1/2}$. Thus $r_0$ is regarded as the tube radius.

Suppose that the vortex tube penetrates the $(x,y)$ plane at the point $(0,\Delta)$. The $x$ and $y$ axes are, respectively, in the streamwise and transverse directions. If the direction of the tube axis is $(\theta, \varphi)$ in spherical coordinates, the streamwise ($u$) and transverse ($v$) components of the circulation flow $u_{\Theta}$ along the $x$ axis are
\begin{subequations}
\begin{equation}
\label{eq2a}
u = \frac{\Delta \cos \theta}{r} u_{\Theta} (r)
\end{equation}
and
\begin{equation}
\label{eq2b}
v = \frac{x \cos \theta}{r} u_{\Theta} (r),
\end{equation}
\end{subequations}
with
\begin{equation}
\label{eq3}
r^2 = x^2           ( 1 - \sin ^2 \theta \cos ^2 \varphi ) +
      \Delta ^2 ( 1 - \sin ^2 \theta \sin ^2 \varphi ) +
      2x \Delta \sin ^2 \theta \sin \varphi \cos \varphi.
\end{equation}
Likewise, for the radial inflow $u_r$ of the strain field, the streamwise and transverse components are
\begin{subequations}
\begin{equation}
\label{eq4a}
u = \frac
    {x ( 1 - \sin ^2 \theta \cos ^2 \varphi ) + \Delta \sin ^2 \theta \sin \varphi \cos \varphi }
    {r} 
    u_r (r) 
\end{equation}
and
\begin{equation}
\label{eq4b}
v = - \frac
{x \sin ^2 \theta \sin \varphi \cos \varphi + \Delta ( 1 - \sin ^2 \theta \sin ^2 \varphi )}
{r} 
u_r (r).
\end{equation}
\end{subequations}
If the vortex tube passes close to the probe ($\Delta \alt r_0$) and the tube is not heavily inclined ($\theta \simeq 0$), the transverse velocity is dominated by the circulation flow. The streamwise velocity is dominated by the radial inflow. This situation is of particular interest. If $\Delta \gg r_0$, the tube signal is weak. If $\theta \gg 0$, the transverse velocity is dominated by the radial inflow, which is of large scale and does not contribute to the small-scale intermittency.

The velocity profiles of vortex tubes with $\Delta \alt r_0$ and $\theta \simeq 0$ are nearly the same \cite{mtk99,b96}. However, as implied by the first term of Eq. (\ref{eq3}), the radius $r_x$ of a tube observed along the streamwise direction is different from its intrinsic radius $r_0$ as $r_x \simeq r_0 / (1- \sin ^2 \theta \cos ^2 \varphi )^{1/2}$. Thus, even if the intrinsic radii of the vortex tubes are the same, they contribute to a range of the scales.

\section{RESULTS AND DISCUSSION}
\label{s4}

Hereafter we describe our experimental results. Except for Fig.~\ref{f6}, the following diagrams present the results only for the experiments 1, 2, 5, 8, and 10 (see Table~\ref{t1}). We nevertheless obtained consistent results for the other experiments.

\subsection{Velocity profile of vortex tube}
\label{s4a}

The streamwise-velocity increment $\delta u(\delta x)$ = $u(x+ \delta x) - u(x)$ or the transverse-velocity increment $\delta v(\delta x)$ = $v(x+ \delta x) - v(x)$ is enhanced in an intermittency event. We obtain velocity patterns typical of the intermittency events by averaging signals centered at the position where the transverse-velocity increment is enhanced over its root-mean-square value by more than a factor of 3, i.e., $\vert \delta v \vert > 3 \langle \delta v^2 \rangle ^{1/2}$. The scale $\delta x$ is set to be the spatial resolution $U / f_c$. Such an event is detected 2.1--5.9 times per the integral length. When the increment is negative, we invert the sign of the $v$ signal. 

The numerical factor of 3 in our above criterion comes from a compromise. If the factor is larger, the total number of the events is smaller, and hence statistical uncertainty in the velocity patterns is more significant. If the factor is smaller, the velocity patterns are less representative of intermittency. It was nevertheless ascertained that our following results depend only weakly on the choice of the numerical factor if it is in the range 2--4. 

The results are shown in Fig.~\ref{f1} (solid lines). The $u$ pattern is shown separately for $\delta u > 0$ and $\delta u \le 0$ at $x = 0$ (designated, respectively, as $u^+$ and $u^-$). For reference, we show velocity profiles (\ref{eq2b}) and (\ref{eq4a}) of a Burgers vortex with $\Delta = 0$ and $\theta = 0$ (dotted lines). The tube radius $r_0$ was determined so as to reproduce the $v$ pattern for each of the experiments.

%%%% Place Figure 1 %%%%

The $v$ pattern of grid turbulence is close to the profile of a Burgers vortex \cite{n97,cg99}. Vortex tubes are surely responsible for small-scale intermittency in the transverse velocity. The radii $r_0$ of the vortex tubes are estimated to be several of the Kolmogorov length $\eta$, $r_0$ $\simeq$ (6--7) $\eta$, regardless of the Reynolds number Re$_{\lambda}$ \cite{j93,tmi99,b96}. Since these estimates are well above the $\delta x$ values, they should represent typical radii of vortex tubes. 

Especially when the Reynolds number is high, the $v$ pattern is extended as compared with the profile of a Burgers vortex. There should be additional contributions from vortex sheets, for which the velocity signal exhibits some kind of step, and also from vortex tubes with $\theta \gg 0$.

The $u^{\pm}$ patterns of grid turbulence appear to be dominated by the circulation flows $u_{\Theta}$ of vortex tubes passing the probe at some distances $\Delta > 0$ [Eq. (\ref{eq2a})]. There is no significance of the radial inflow $u_r$ of the strain field [Eq. (\ref{eq4a})], except that the $u^-$ pattern has a larger amplitude and is detected 1.3--1.5 times more frequently than the $u^+$ pattern. A vortex tube is not always oriented to the stretching direction \cite{vm91,j93,t92}.

The amplitude of the velocity pattern is normalized by the Kolmogorov velocity $u_K$. This normalized amplitude is nearly the same for different Reynolds numbers. If normalized by the root-mean-square velocity fluctuation $\langle u^2 \rangle ^{1/2}$ or $\langle v^2 \rangle ^{1/2}$, the amplitude is smaller at a higher Reynolds number. The typical strength of vortex tubes appears to scale with the Kolmogorov velocity rather than the root-mean-square velocity fluctuation, although the present result is biased toward strong tubes.

The $u^{\pm}$ patterns are embedded in one-sided excursions, $u \simeq u_K \simeq$ a few tenths of $\langle u^2 \rangle ^{1/2}$ $>$ 0, which extends to $x$ $\simeq$ $\pm L$. This could be due to an artifact of our measurement or analysis. For example, the streamwise velocity has a fluctuation of the order $\langle u^2 \rangle ^{1/2}$ over the scale $L$. Thus the speed at which a vortex tube passes the probe is higher or lower than the mean wind speed. The tube radius estimated with the frozen-eddy hypothesis is accordingly smaller or larger than its intrinsic $r_x$ value. Since the scale $\delta x = U/f_c$ is at around the typical tube radius $r_0$, vortex tubes with $\theta > 0$ and hence $r_x > \delta x$ are more numerous than those with $r_x < \delta x$. Fast-moving vortex tubes with $r_x > \delta x$ are captured in our conditional averaging for the scale $\delta x$ and could cause the observed $u$ excursions. Alternatively, large-scale flows that are responsible for the excursions could preferentially generate vortex tubes.

For the larger scales $\delta x \alt \lambda$, our conditional averaging yields nearly the same patterns as those in Fig.~\ref{f1}. Velocity signals at large scales reflect vortex tubes that are inclined to the mean-wind direction (Sec.~\ref{s3}). Large-radius vortex tubes could be also important. With increasing the scale still more, the background flow is increasingly predominant (Sec.~\ref{s4c}). It is increasingly difficult to capture vortex tubes.

%%%% Place Figures 2 and 3 %%%%

Velocity patterns are extracted also for enhancements of the streamwise-velocity increment $\delta u(\delta x)$ at the scale $\delta x = U/f_c$. The results for $\delta u > +3 \langle \delta u^2 \rangle ^{1/2}$ are shown in Fig.~\ref{f2}, while those for $\delta u < -3 \langle \delta u^2 \rangle ^{1/2}$ are shown in Fig.~\ref{f3}. The $v$ pattern was obtained by conditioning over the sign of the local slope. Since the $v$ pattern is close to the profile of a Burgers vortex, intermittency in the streamwise velocity is also attributable to vortex tubes. The tube radii are several of the Kolmogorov length. On the other hand, the $u$ pattern is explained by the circulation flow $u_{\Theta}$ passing the probe at some distance $\Delta > 0$. The radial inflow $u_r$ is also important to the $u$ pattern for the event $\delta u < -3 \langle \delta u^2 \rangle ^{1/2}$, which is detected 1.6--2.5 times more frequently than the event $\delta u > +3 \langle \delta u^2 \rangle ^{1/2}$ (see also Sec.~\ref{s4d}).

\subsection{Spatial distribution of vortex tubes}
\label{s4b}

To study the spatial distribution of vortex tubes, we study the probability density distribution of an interval $\delta x'$ between successive enhancements of the transverse-velocity increment, $\vert \delta v \vert > 3 \langle \delta v^2 \rangle ^{1/2}$. Here $\delta v (\delta x) = v(x+\delta x)-v(x)$ is computed at the scale $\delta x = U/f_c$.

%%%% Place Figure 4 %%%%

The results are shown in Fig.~\ref{f4} (solid lines). The probability density distribution has an exponential tail that appears linear on the semi-log plot (dotted lines). This exponential distribution of the interval is characteristic of random, independent, and rare events \cite{cdc95}. However, at a small interval that is several times the Kolmogorov scale $\eta$, the distribution has a pronounced peak. Vortex tubes tend to cluster together \cite{cdc95,lvmb00,
cg99}.

With increasing the Reynolds number Re$_{\lambda}$, the probability density distribution becomes to have a more pronounced peak. The clustering of vortex tubes becomes more significant. Figures \ref{f1}--\ref{f3} actually show that the intermittency patterns at high Reynolds numbers are somewhat uneven.

\subsection{Flatness factor}
\label{s4c}

For velocity increments $\delta u(\delta x)$ = $u(x+\delta x)-u(x)$ and $\delta v(\delta x)$ = $v(x+\delta x)-v(x)$, we compute the flatness factors at each of the scales $\delta x$:
\begin{equation}
F_{\delta u} = \frac{ \langle \delta u^4 \rangle }{ \langle \delta u^2 \rangle ^2}
\quad {\rm and} \quad
F_{\delta v} = \frac{ \langle \delta v^4 \rangle }{ \langle \delta v^2 \rangle ^2}.
\end{equation}
If the probability density distribution of the velocity increment is Gaussian, the flatness factor is equal to 3. If the distribution has a more pronounced tail than a Gaussian distribution, the flatness factor is larger than 3. Since the transverse-velocity pattern averaged for enhancements of velocity increments is close to the profile of a Burgers vortex (Sec.~\ref{s4a}), the flatness factor serves as a measure of the relative importance between vortex tubes and the background flow. The latter is a superposition of random and independent motions, for which the velocity increment is expected to follow a Gaussian distribution \cite{mtk99,mkfnt99,mthk02}.

%%%% Place Figure 5 %%%%

Figure~\ref{f5} shows the flatness factor for the transverse-velocity increment $\delta v$ (solid lines). At large scales, the flatness factor is equal to the Gaussian value of 3. This is due to predominance of the background flow. As the scale is decreased from the integral length $L$, the flatness factor begins to increase. This is due to increasing contribution of vortex tubes. They affect velocity increments at scales around the observed sizes $r_x$, which could be as large as the tube lengths that are of the order of the integral length \cite{vm91,j93,tmi99,d91,cdc95,lvmb00,f95,sa97,s99}. Around the Taylor microscale $\lambda$, the increase of the flatness factor becomes significant.

Figure~\ref{f5} also shows the flatness factor for the streamwise-velocity increment $\delta u$, which is less enhanced than that for the transverse-velocity increment $\delta v$ \cite{vm91,j93,azs96}. Judging from Figs.~\ref{f2} and \ref{f3}, the streamwise flatness factor reflects the energetical importance of vortex tubes as in the case of the transverse flatness factor. However, the amplitude of the $u$ pattern of a vortex tube tends to be less pronounced than that of the $v$ pattern. This is because the circulation flow $u_{\Theta}$ of a vortex tube contributes to the streamwise velocity only if its axis passes at some distance from the probe, $\Delta > 0$ [Eq. (\ref{eq2a})].

%%%% Place Figure 6 %%%%

Figure~\ref{f6}(a) shows the flatness factors for increments of the streamwise (filled symbols) and transverse (open symbols) velocities at the fixed scales $\delta x = 10 \eta$ (circles) and $\lambda$ (squares). The scale $\delta x = 10 \eta$ is roughly the minimum scale resolved in all the measurements. With increasing the Reynolds number Re$_{\lambda}$, the flatness factors increase. Vortex tubes are increasingly predominant over the background flow at a higher Reynolds number. The increase of the flatness factors is more significant at the smaller scale $\delta x = 10 \eta$, where the background flow is less important. The increase is also more significant for the transverse-velocity increment \cite{j93}, which is more sensitive to vortex tubes than the streamwise-velocity increment.

\subsection{Skewness factor}
\label{s4d}

The skewness factor for the streamwise-velocity increment $\delta u(\delta x)= u(x+\delta x)-u(x)$ is also studied at each of the scales $\delta x$. The definition of the skewness factor is
\begin{equation}
S_u = \frac{ \langle \delta u^3 \rangle }{ \langle \delta u^2 \rangle ^{3/2}}.
\end{equation}
This quantity characterizes the degree of asymmetry. Its negative value implies that the probability density distribution has a pronounced tail extending to the negative $\delta u$.

%%%% Place Figure 7 %%%%

The results are shown in Fig.~\ref{f7} (solid lines). At large scales that are more than a few times greater than the rod spacing of the grid $M$, the skewness factor is equal to the Gaussian value of 0. As the scale is decreased, the skewness factor begins to decrease. This decrease reflects the energy cascade from large to small scales \cite{f95,sa97}. Around the Taylor microscale $\lambda$, the decrease becomes significant. This negative enhancement of the velocity increment is due to the strain fields associated with vortex tubes (Sec.~\ref{s4a}). 

We find no significant dependence of the skewness factor at a fixed scale on the Reynolds number Re$_{\lambda}$. This is because our experiments do not cover a sufficiently wide Re$_{\lambda}$ range. It is known that the skewness factor of the velocity derivative $\langle ( \partial u / \partial x )^3 \rangle / \langle ( \partial u / \partial x )^2 \rangle ^{3/2}$ decreases slowly with increasing the Reynolds number \cite{sa97}.

\subsection{Streamwise-transverse correlation}

Since our conditional averaging revealed that the circulation flow of a vortex tube tends to contribute simultaneously to the streamwise and transverse velocities (Sec.~\ref{s4a}), we study the correlation between velocity increments of these two components \cite{mkfnt99}:
\begin{equation}
C_{\delta u \delta v} 
   = \frac{  \langle \delta u^2 \delta v^2 \rangle
           - \langle \delta u^2 \rangle \langle \delta v^2 \rangle}
          {( \langle \delta u^4 \rangle - \langle \delta u^2 \rangle ^2 ) ^{1/2}
           ( \langle \delta v^4 \rangle - \langle \delta v^2 \rangle ^2 ) ^{1/2}}.
\end{equation}
The value of $C_{\delta u \delta v}$ ranges from 0, if there is no correlation, to $\pm 1$, if there is a linear correlation.

%%%% Place Figure 7 %%%%

The results are shown in Fig.~\ref{f8} (solid lines). At large scales that are more than a few times greater than the rod spacing of the grid $M$, the correlation is absent. As the scale is decreased, the correlation coefficient begins to increase. This is because variations of the streamwise and transverse velocities tend to be enhanced simultaneously at the position of a strong large-scale motion. The increase of the correlation coefficient becomes significant around the integral length $L$ and becomes still more significant around the Taylor microscale $\lambda$. This is due to increasing contribution of vortex tubes.

Figure~\ref{f6}(b) shows the streamwise-transverse correlation at the fixed scales $\delta x = 10 \eta$ (circles) and $\lambda$ (squares). With increasing the Reynolds number Re$_{\lambda}$, the correlation becomes significant especially at the smaller scale $\delta x = 10 \eta$.

\section{Conclusion}
\label{s5}

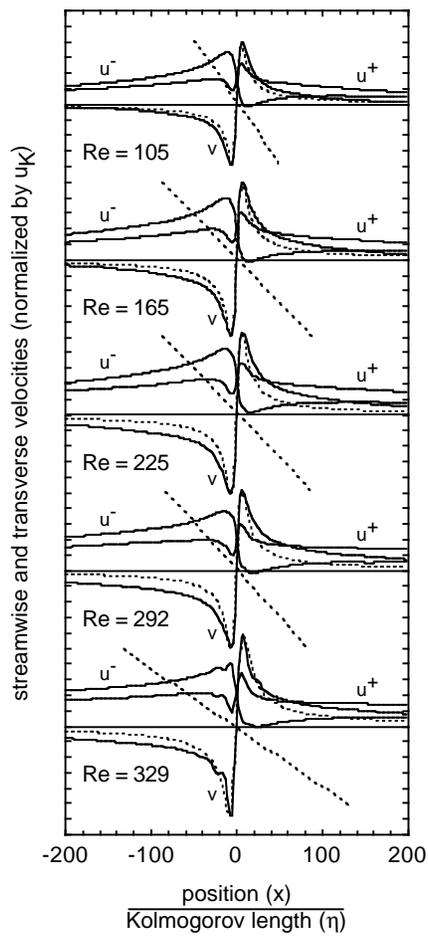
\begin{figure}
\caption{\label{f1} Conditional averages of the streamwise ($u$) and transverse ($v$) velocities for $\vert \delta v \vert > 3 \langle \delta v^2 \rangle ^{1/2}$ at the scale $\delta x$ = $U / f_c$ in the experiments 1, 2, 5, 8, and 10 (from top to bottom). The abscissa is the spatial position $x$ normalized by the Kolmogorov length $\eta$. On the ordinate, the unit scale corresponds to the Kolmogorov velocity $u_K$. We show the streamwise velocity separately for $\delta u > 0$ ($u^+$) and $\delta u \le 0$ ($u^-$). The peak amplitudes of the $v$ patterns are 0.179, 0.179, 0.315, 0.306, and 0.449 m~s$^{-1}$. The values of $3 \langle \delta v^2 \rangle ^{1/2}$ are 0.150, 0.103, 0.270, 0.267, and 0.430 m s$^{-1}$. The detection rates of the $u^+$ event per the integral length are 0.9, 2.1, 2.1, 2.4, and 2.5. Those of the $u^-$ event are 1.2, 2.8, 3.0, 3.4, and 3.4. The profiles of Burgers vortices are shown with dotted lines. Their radii $r_0$ are $5.5 \eta$, $5.9 \eta$, $5.8 \eta$, $5.8 \eta$ and $7.2 \eta$.}
\end{figure}

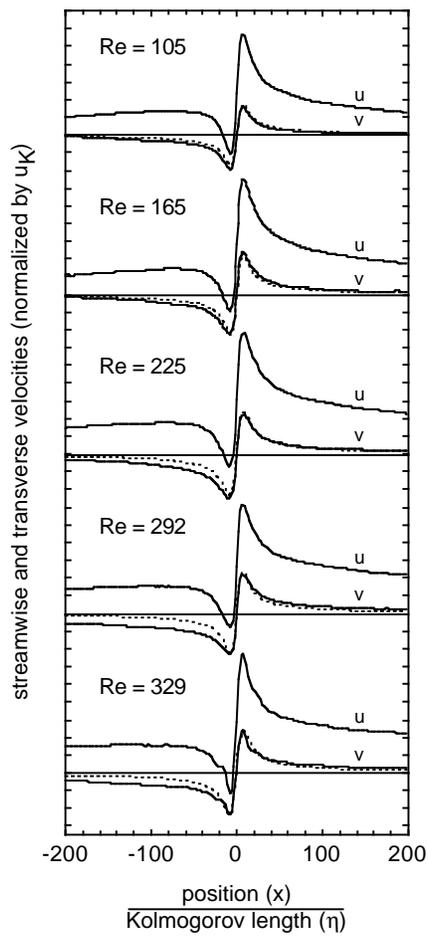
\begin{figure}
\caption{\label{f2} Conditional averages of the streamwise ($u$) and transverse ($v$) velocities for $\delta u > +3 \langle \delta u^2 \rangle ^{1/2}$ at the scale $\delta x$ = $U / f_c$ in the experiments 1, 2, 5, 8, and 10 (from top to bottom). The abscissa is the spatial position $x$ normalized by the Kolmogorov length $\eta$. On the ordinate, the unit scale corresponds to the Kolmogorov velocity $u_K$. The peak amplitudes of the $u$ patterns are 0.256, 0.233, 0.407, 0.364, and 0.516 m~s$^{-1}$. The values of $3 \langle \delta u^2 \rangle ^{1/2}$ are 0.124, 0.071, 0.190, 0.187, and 0.299 m~s$^{-1}$. The detection rates per the integral length are 0.5, 1.1, 1.3, 1.6 and 1.8. The profiles of Burgers vortices are shown with dotted lines. Their radii $r_0$ are $5.5 \eta$, $6.5 \eta$, $7.7 \eta$, $6.7 \eta$, and $7.2 \eta$.}
\end{figure}

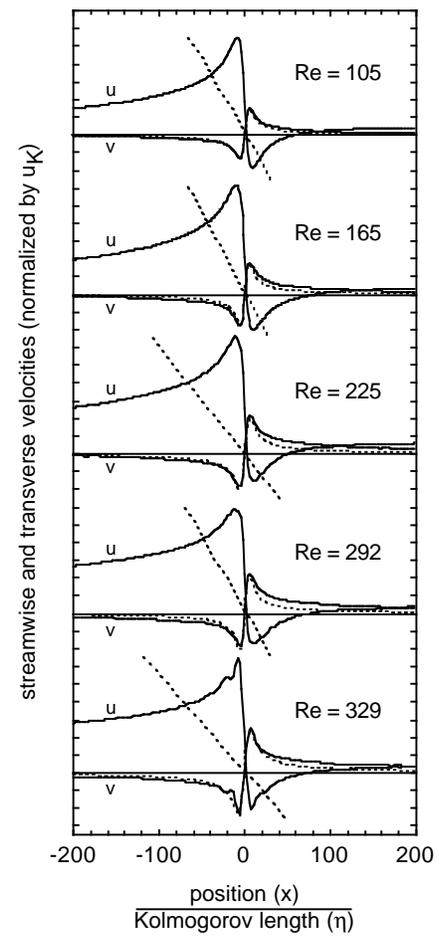
\begin{figure}
\caption{\label{f3} Conditional averages of the streamwise ($u$) and transverse ($v$) velocities for $\delta u$ $<$ $-3 \langle \delta u^2 \rangle ^{1/2}$ at the scale $\delta x$ = $U / f_c$ in the experiments 1, 2, 5, 8, and 10 (from top to bottom). The abscissa is the spatial position $x$ normalized by the Kolmogorov length $\eta$. On the ordinate, the unit scale corresponds to the Kolmogorov velocity $u_K$. The peak amplitudes of the $u$ patterns are 0.246, 0.221, 0.396, 0.349, and 0.491 m~s$^{-1}$. The detection rates per the integral length are 1.2, 2.7, 2.7, 3.0, and 3.0. The profiles of Burgers vortices are shown with dotted lines. Their radii $r_0$ are $5.5 \eta$, $4.7 \eta$, $5.8 \eta$, $4.8 \eta$, and $6.0 \eta$.}
\end{figure}

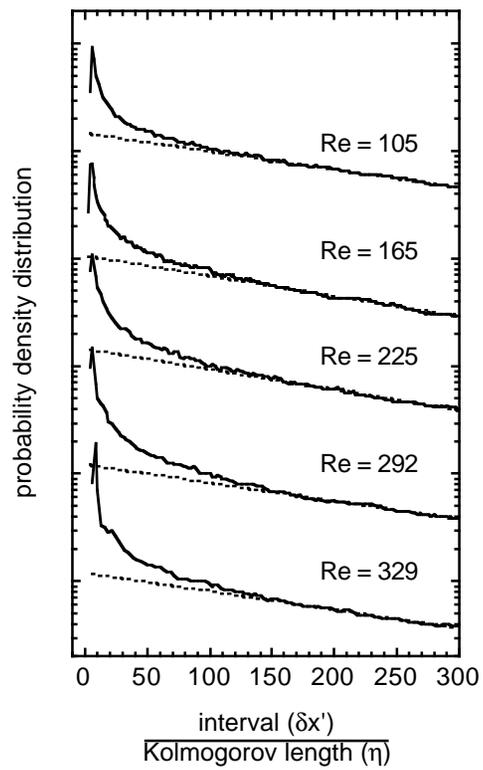
\begin{figure}
\caption{\label{f4} Probability density distribution of an interval $\delta x'$ between successive enhancements of the transverse-velocity increment $\vert \delta v \vert > 3 \langle \delta v^2 \rangle ^{1/2}$ at the scale $\delta x = U/f_c$ in the experiments 1, 2, 5, 8, and 10 (from top to bottom). We vertically shift the probability density distribution by a factor 10. The abscissa is the interval $\delta x'$ normalized by the Kolmogorov length $\eta$. The dotted lines denote the results of exponential fit at large intervals.}
\end{figure}

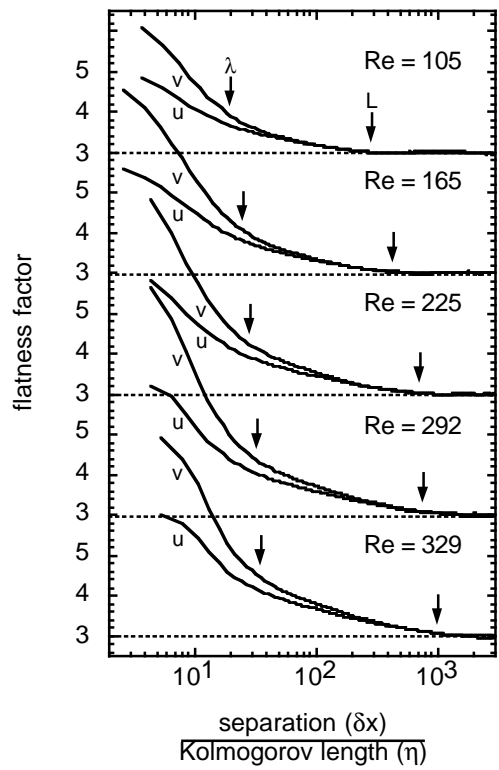
\begin{figure}
\caption{\label{f5} Flatness factors for increments of the streamwise ($u$) and transverse ($v$) velocities in the experiments 1, 2, 5, 8, and 10 (from top to bottom). The abscissa is the scale $\delta x$ normalized by the Kolmogorov length $\eta$. The arrows denote the integral length $L$ and the Taylor microscale $\lambda$. The dotted lines denote the Gaussian value of 3.}
\end{figure}

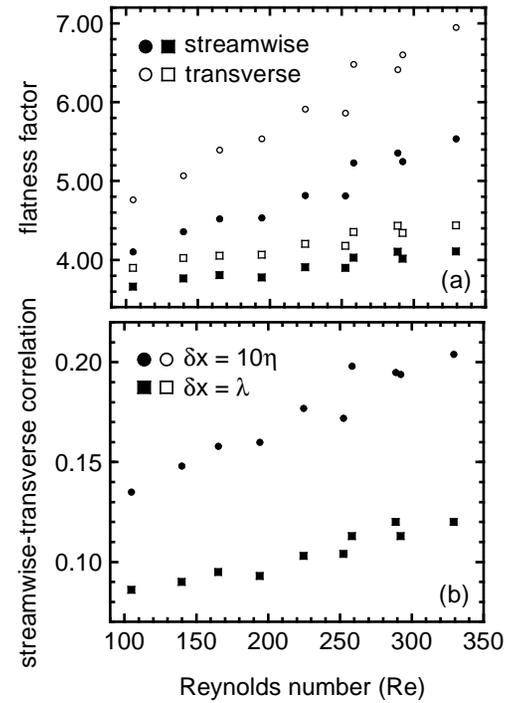
\begin{figure}
\caption{\label{f6} (a) Flatness factors for increments of the streamwise (filled symbols) and transverse (open symbols) velocities. (b) Correlation between increments of the streamwise and transverse velocities. The abscissa is the Reynolds number Re$_{\lambda}$. The circles denote the scale $\delta x = 10 \eta$, while the squares denote the scale $\delta x = \lambda$.}
\end{figure}

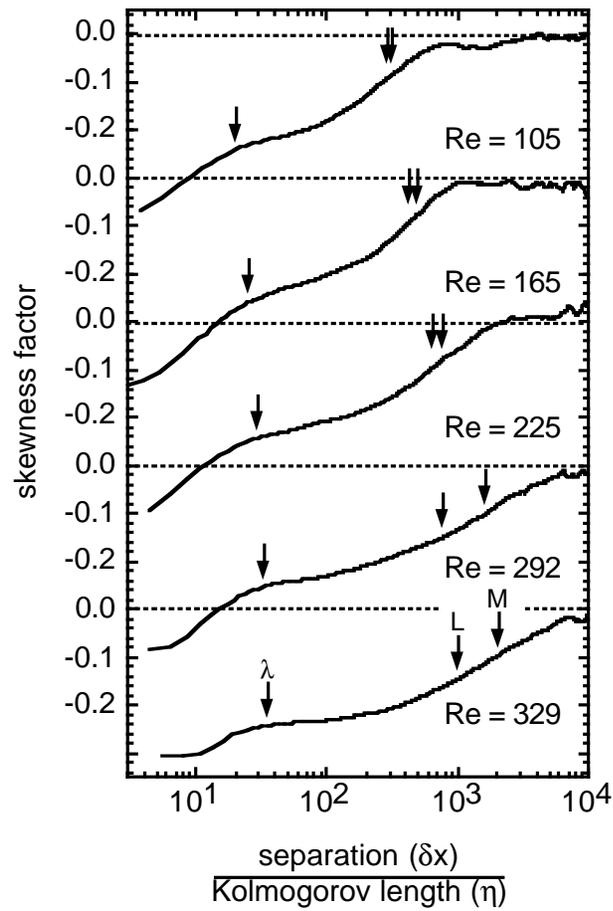
\begin{figure}
\caption{\label{f7} Skewness factor for the streamwise-velocity increment in the experiments 1, 2, 5, 8, and 10 (from top to bottom). The abscissa is the scale $\delta x$ normalized by the Kolmogorov length $\eta$. The arrows denote the rod spacing of the grid $M$, the integral length $L$, and the Taylor microscale $\lambda$. The dotted lines denote the Gaussian value of 0.}
\end{figure}

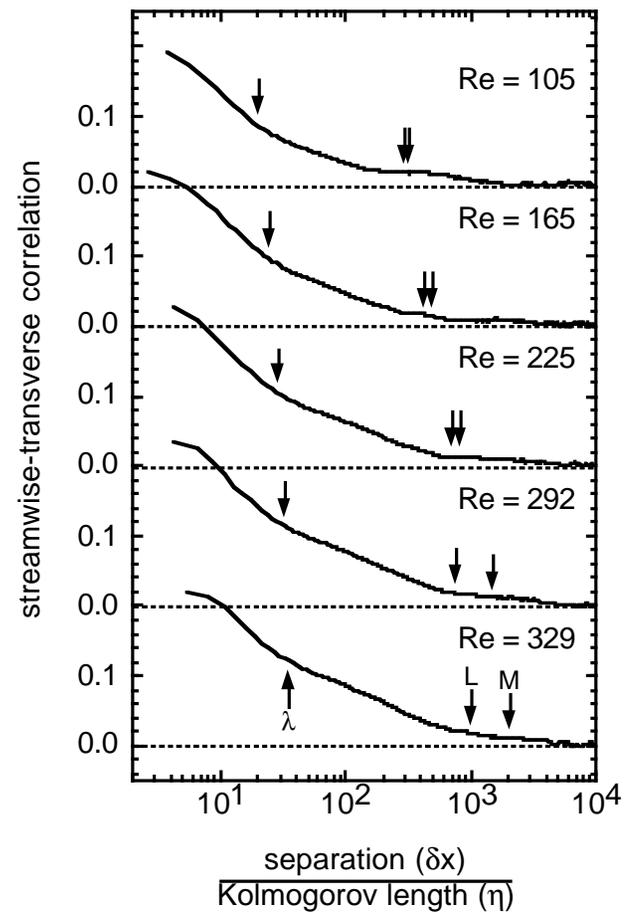
\begin{figure}
\caption{\label{f8} Correlation between increments of the streamwise and transverse velocities in the experiments 1, 2, 5, 8, and 10 (from top to bottom). The abscissa is the scale $\delta x$ normalized by the Kolmogorov length $\eta$. The arrows denote the rod spacing of the grid $M$, the integral length $L$, and the Taylor microscale $\lambda$. The dotted lines denote the non-correlation value of 0. }
\end{figure}

The streamwise ($u$) and transverse ($v$) velocities were measured simultaneously in isotropic grid turbulence at Reynolds numbers Re$_{\lambda} \simeq 110$--330. First, we used conditional averaging to extract typical intermittency patterns (Figs.~\ref{f1}--\ref{f3}). Since the $v$ pattern is close to the velocity profile of a Burgers vortex, the intermittency is attributable to vortex tubes. The tube radius is several times the Kolmogorov length regardless of the Reynolds number. Since the $u$ pattern is not always dominated by the strain field, it was concluded that a vortex tube is not always oriented to the stretching direction. Second, we used the distribution of an interval between successive enhancements of a small-scale velocity increment to study the spatial distribution of vortex tubes (Fig.~\ref{f4}). It was found that vortex tubes tend to cluster together. This tendency becomes significant with the Reynolds number. Third, we used the flatness, skewness, and streamwise-transverse correlation of velocity increments to study the energetical contribution of vortex tubes as a function of the scale (Figs.~\ref{f5}--\ref{f8}). The contribution of vortex tubes is discernible below the integral length, which is comparable to the tube lengths, and it is significant below the Taylor microscale. At a fixed scale, the contribution becomes significant with the Reynolds number.

Some of the above results had been reported in past studies. However, they are either numerical simulations with low Reynolds numbers, Re$_{\lambda} \alt 200$ \cite{vm91,j93,tmi99}, laboratory experiments of isotropic turbulence with low Reynolds numbers, Re$_{\lambda} \alt 100$ \cite{t92}, or laboratory experiments of anisotropic turbulence \cite{cdc95,lvmb00,n97,cg99,b96,azs96}. We are the first to study vortex tubes in isotropic turbulence at relatively high Reynolds numbers, Re$_{\lambda} \simeq 110$--330.

Since some tube parameters change as the Reynolds number, experimental studies of them at the higher Reynolds numbers are desirable. It is difficult to achieve Re$_{\lambda} \gg 300$ in ordinary grid turbulence. Other flow configurations such as a boundary layer and a jet are of interest.

The velocity patterns of vortex tubes studied here were biased toward strong tubes because they were obtained by averaging for enhancements of a velocity increment. If velocity patterns averaged for vortex tubes with various strengths were available, they would be more representative. Then it would be possible to confirm possible interesting properties of vortex tubes, e.g., dependence of their typical strengths on the Reynolds number. Mouri and Takaoka \cite{mt02} proposed to identify vortex tubes as local maxima on the scale-space plot of wavelet transforms of the velocity data (see also Ref. \cite{mtk99}). The application of such a technique to turbulence data with a range of the Reynolds number is desirable.

\begin{acknowledgments}
This research has been supported in part by the Japanese Ministry of Education, Science, and Culture under grant (B2) 14340138. The authors are grateful to M. Takaoka for interesting discussion and to T. Umezawa for assistance of our experiments.
\end{acknowledgments}

%%%% References %%%%

%%%% TABLE 1 %%%%

%\clearpage

\begingroup
\squeezetable
\begin{table}
\caption{\label{t1} 
Summary of experimental conditions and turbulence parameters: mean wind speed $U$, rod spacing of the grid $M$, distance from the grid $d$, sampling frequency $f_s$, kinematic viscosity $\nu$ that reflects the air temperature, flatness factors of the velocity fluctuations $F_u$ and $F_v$, mean energy dissipation rate $\langle \varepsilon \rangle$, root-mean-square velocity fluctuations $\langle u^2 \rangle ^{1/2}$ and $\langle v^2 \rangle ^{1/2}$, Kolmogorov velocity $u_K$, integral length $L$, Taylor microscale $\lambda$, Kolmogorov length $\eta$, and Reynolds number $Re_{\lambda}$. Velocity derivatives were obtained as, e.g., $\partial u / \partial x$ = $[ 8 u(x+ \delta x)- 8 u(x- \delta x)-u(x+ 2 \delta x)+u(x- 2 \delta x)]/ (12 \delta x)$ with $\delta x = U/f_s$. The experiments 1 and 4 were made in the 0.8 $\times$ 0.8 $\times$ 3 m tunnel, while the other experiments were made in the 3 $\times$ 2 $\times$ 18 m tunnel.}

\begin{ruledtabular}
\begin{tabular}{llcccccccccc}
                                                                                        &                  & 1      & 2      & 3      & 4      & 5      & 6      & 7      & 8      & 9      & 10     \\ 
\hline
$U$                                                                                     & m s$^{-1}$       & 4.76   & 4.21   & 4.31   & 9.37   & 8.33   & 8.82   & 12.4   & 12.6   & 16.3   & 17.0   \\
$M$                                                                                     & cm               & 10     & 20     & 40     & 10     & 20     & 40     & 20     & 40     & 20     & 40     \\
$d$                                                                                     & m                & 2.0    & 3.5    & 6.5    & 2.0    & 3.5    & 8.0    & 3.5    & 8.0    & 4.0    & 8.0    \\
$f_s$                                                                                   & kHz              & 8      & 8      & 8      & 16     & 16     & 16     & 24     & 24     & 32     & 32     \\
$\nu$                                                                                   & cm$^2$ s$^{-1}$  & 0.145  & 0.143  & 0.143  & 0.160  & 0.145  & 0.149  & 0.146  & 0.146  & 0.148  & 0.150  \\
\hline
$F_u = \langle u^4 \rangle / \langle u^2 \rangle ^2$                                    &                  & 3.00   & 3.00   & 2.90   & 2.99   & 3.01   & 2.95   & 3.03   & 2.95   & 3.03   & 2.96   \\
$F_v = \langle v^4 \rangle / \langle v^2 \rangle ^2$                                    &                  & 2.95   & 3.01   & 2.94   & 2.96   & 2.99   & 3.02   & 2.99   & 3.03   & 3.02   & 3.03   \\
$\langle \varepsilon \rangle = 15 \nu \langle (\partial u/ \partial x )^2 \rangle$      & m$^2$ s$^{-3}$   & 0.282  & 0.114  & 0.0507 & 2.17   & 0.890  & 0.289  & 3.20   & 0.866  & 5.71   & 2.26   \\ 
$\langle u^2 \rangle ^{1/2}$                                                            & m s$^{-1}$       & 0.234  & 0.233  & 0.207  & 0.461  & 0.457  & 0.368  & 0.675  & 0.518  & 0.828  & 0.704  \\
$\langle v^2 \rangle ^{1/2}$                                                            & m s$^{-1}$       & 0.193  & 0.227  & 0.201  & 0.404  & 0.446  & 0.347  & 0.666  & 0.505  & 0.819  & 0.682  \\
$u_K = ( \nu \langle \varepsilon \rangle )^{1/4}$                                       & m s$^{-1}$       & 0.0450 & 0.0358 & 0.0292 & 0.0768 & 0.0599 & 0.0455 & 0.0827 & 0.0596 & 0.0959 & 0.0763 \\
$L = \int \langle u(x+\delta x) u(x) \rangle / \langle u^2 \rangle d \delta x$          & cm               & 9.46   & 17.3   & 17.8   & 11.5   & 17.2   & 17.9   & 18.4   & 19.1   & 18.5   & 19.9   \\
$\lambda = [ \langle u^2 \rangle / \langle (\partial u/ \partial x )^2 \rangle ]^{1/2}$ & cm               & 0.650  & 1.01   & 1.34   & 0.485  & 0.713  & 1.02   & 0.559  & 0.824  & 0.516  & 0.702  \\
$\eta = (\nu ^3 / \langle \varepsilon \rangle )^{1/4}$                                  & cm               & 0.0322 & 0.0400 & 0.0490 & 0.0208 & 0.0242 & 0.0327 & 0.0177 & 0.0245 & 0.0154 & 0.0197 \\
Re$_{\lambda} = \langle u^2 \rangle ^{1/2} \lambda / \nu$                               &                  & 105    & 165    & 194    & 140    & 225    & 252    & 258    & 292    & 289    & 329    \\

\end{tabular}
\end{ruledtabular}
\end{table}
\endgroup
%%%% Figure Captions %%%%

\clearpage

\end{document}